\documentclass{acaces}

\newcommand*{\dquote}[1]{``#1"}

\begin{document}

\title{
\begin{minipage}[t]{0.9\textwidth}
\flushleft
Perfrewrite\\
Program Complexity Analysis via Source Code Instrumentation
\end{minipage}
}

\author{
Michael~Kruse\addressnum{1}\comma\extranum{1},
}

\address{1}{
INRIA Saclay {\^I}le-de-France,
Parc Orsay Universit{\'e},
4 rue Jacques Monod,\\
91893 Orsay cedex,
France
}

\extra{1}{E-mail: michael.kruse@inria.fr}

\pagestyle{empty}

\begin{abstract}
Most program profiling methods output the execution time of one specific program execution, but not its computational complexity class in terms of the big-O notation. Perfrewrite is a tool based on LLVM's Clang compiler to rewrite a program such that it tracks semantic information while the program executes and uses it to guess memory usage, communication and computational complexity. While source code instrumentation is a standard technique for profiling, using it for deriving formulas is an uncommon approach.
\end{abstract}


\section{Motivation}

Several methods are common in order to deduce the performance characteristics of a program, usually used to identify a program's bottleneck that is most worthy to optimize. We are specially interested in HPC applications used by physicists. Our case study are the programs tmLQCD~\cite{tmlqcd} and DD-HMC~\cite{ddhmc}, both implementations of the lattice quantum chromodynamics (LQCD) simulation. The goal is to automatically derive the execution time in terms of the size of the input field.

The usual approach for program complexity analysis is done statically on the source code with tools such as \emph{PIPS}~\cite{pips}. Unfortunately, challenges like pointer arithmetic make it near impossible for such tools to process the given programs. Another idea is to use modelling program such as \emph{PAMELA}~\cite{pamela}, but it requires to rewrite a program in its domain specific language and still does not support language constructs such as pointers.

Profiling with tools like \emph{gprof}, \emph{oprofile}, \emph{Tau} and many more is the most mature technique, but do not return the complexity in big-O notation. This drawback can be coped with by running the program multiple times with different inputs and let a statistics tool (like \emph{GNU R}) analyse it. To get a meaningful result also large input sizes have to be sampled. In case of shared cluster systems probably someone will complain about eating computation time just to find out how long a program takes to execute.

\section{Approach}

The main idea is to replace all relevant types in a C program by custom C++ classes. For instance, the \texttt{double} type gets replaced with a custom \texttt{Double}\footnote{Notice the change of capitalization} class. Then, we rely on C++ operator overloading to call our method whenever an arithmetic operation is to be executed. In addition to just simply execute the operation, it increases a global \emph{FLOP} counter. Similarly, \texttt{malloc} and \texttt{free} can be substituted to track memory usage as well as calls to MPI to track communication between cluster nodes.

By itself this does nothing more than standard profiling does. In addition, also loops are annotated using preprocessor macros. If the number of loop iterations is dependent on the program's input size, say $n$, it will multiply the counters in the loop body by $n$ and divide by the number of loop iterations it actually executes in its configuration. In order to make this work, integral types are also replaced by custom C++ classes such that they carry annotations of their values' origin. The associated semantic information are the actual integer value, the value expressed as a term (a symbolic formula of the input size), and a typical integer value of a typical large input size.

For example, a program has two inputs N and M of size $n$ and $m$. We execute the program using a configuration $n=8$ and $m=4$, a very small problem size. The symbolic representations are $n_\mathrm{term}=\dquote{n}$ and $m_\mathrm{term}=\dquote{m}$ respectively. When the values are used in an arithmetic operation, for instance they are multiplied, the annotated value becomes $nm_\mathrm{term}=\dquote{n \cdot m}$. The annotated large input size might be $n_\mathrm{large}=256$ and $m_\mathrm{large}=128$. The annotation after the multiplication is $nm_\mathrm{large}=32768$. It is used whenever two values are compared since the formula representation has no total order, assuming that the large problem case is more relevant. One use is the update of the peak memory usage during the program's execution, avoiding a gigantic step function. In short, we execute the program for a small problem size while interpolating its characteristics for a large problem size.

As most profiling tools do, we log the begin and return of every function. This will allow to create a call tree with stats for every function.

\begin{figure}[htb]
\begin{tabular}{p{0.45\linewidth}p{0.55\linewidth}}%
\scriptsize\begin{verbatim}void execute(int n) {
  double *field = 
    malloc(n * sizeof(*field)); 
  double localSum = 0;
  for (int i = 0; i < n; ++i) 
    localSum += field[i];
  double globalSum;
  if (n > 128)
  	MPI_Allreduce(&localSum, &globalSum, 
  	  1, MPI_DOUBLE, MPI_SUM, MPI_COMM_WORLD);
  free(field);
}\end{verbatim}&%
\scriptsize\begin{verbatim}void execute(Num n) {ENTERFUNCTION
  DynamicMem<Double> field = 
    perf_malloc<Double>(n * sizeof(*field));
  Double localSum = 0;
  LOOP(n) for (int i = 0; i < n; ++i) ITERATION 
    localSum += field[i];
  Double globalSum;
  if (n > 128)
    MPI_Allreduce(&localSum, &globalSum, 
      1, MPI_DOUBLE, MPI_SUM, MPI_COMM_WORLD);
  free(field);
EXITFUNCTION}\end{verbatim}%
\end{tabular}%
\caption{An example program before (left) and after instrumentation (right)}
\label{exrewrite}
\end{figure}

Of course this approach is not perfect. It works only with polynomial complexities. Loops must have the structure of for-loops with iteration count directly depending on the input. Data-dependent while-loops for instance are not possible, but can be worked-around by making the number of while-iterations an input size. But the advantage is that the techniques does not need to cope with logic unrelated to the complexity. No need to handle pointer arithmetic, polymorphism, the halting problem, etc. because the program is executed and these intractabilities are evaluated naturally.

The implemented approach executes every loop body at most twice. The first iteration may initialize global fields\footnote{of the kind \texttt{if (!g\_field) g\_field = malloc(\dots)}}, the second iteration is assumed to represent all following iterations. The is sufficient for the two case study programs mentioned in the previous section. The program's result will be wrong, but we are not interested in it anyway.

\section{Implementation}

A simple textual insertion and replacement is not sufficient. We want to track memory blocks, therefore we also need to replace pointers to them. Pointers are typed, meaning the pointer replacement class needs to be a template whose variable type declaration syntax is different than a pointer's. To instrument manually is a tedious task to be avoided if possible. In addition, instrumenting only necessary parts of the code will safe us from incompatibilities in unaffected and working program sections. E. g. C++ classes cannot perfectly emulate the behaviour of C pointers. If necessary the user might need to adapt the code manually, like when a loop syntax does not exactly match any implemented patters. This is why the instrumented code must stay readable.

The Clang compiler~\cite{clang} has the necessary facilities for a semantic analysis of C and C++ code. It also retains the source code locations in the internal representation so our tool can locate and replace the parts that have to be replaced. Clang already includes an Objective-C to C++ translator that works similarly. Also, a full C++ compiler already includes facilities to get the return type of an expression involving custom types.
 
\begin{figure}[htb]
\centering
\includegraphics[width=0.6\textwidth]{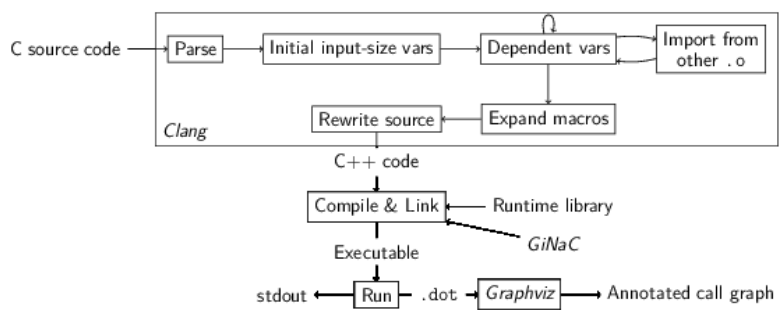}
\caption{Processing pipeline.}
\label{pipeline}
\end{figure}

The principal tool working is shown in Figure~\ref{pipeline}. Intrinsic types are transformed into custom C++ types whenever it is assigned an expression that evaluates to a custom type or the types are incompatible in some context (e.g. when passing by reference). Types may influence themselves, so this is done until a fixpoint is reached. When types types declared \texttt{extern} are changed -- this includes non-static functions -- this change must be propagated to the other source files of the program.  
If the change is part of a macro, the macro has to be expanded before the instrumentation. Clang does not retain the complete information of how a macro was expanded so the complete parse must be repeated after macro expansion.

The instrumented code can now be compiled using any C++ compiler. The code changes require a custom runtime to be linked against it containing the implementation of the replacement classes. The implementation of the symbolic representation is from GiNaC~\cite{ginac}, a computer algebra system using C++ as control language.

\begin{figure}[htb]
\centering
\includegraphics[width=\textwidth]{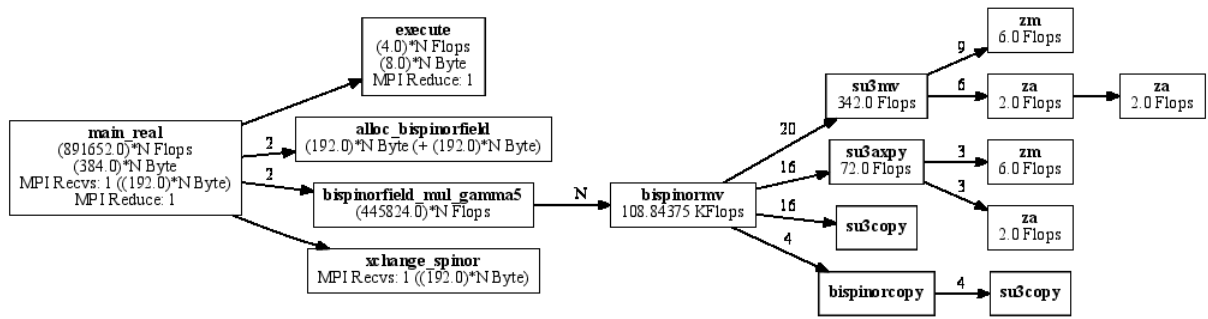}
\caption{Graphviz rendering of an example program.}
\label{perftest_perf}
\end{figure}

When the compilee runs, it writes a call tree into a \texttt{.dot} file that can be processed by Graphviz~\cite{graphviz}. Figure~\ref{perftest_perf} shows such a call tree. For every function call it shows how often it has been called, the number of floating point operations, peak memory usage and MPI calls, all as complexities depending on the input parameters.

\bibliography{guide}

\newcommand{\etalchar}[1]{$^{#1}$}
\begin{thebibliography}{KAC{\etalchar{+}}96}

\bibitem[{AT }]{graphviz}
{AT \& T Labs Research}.
\newblock {Graphviz}.
\newblock \url{http://graphviz.org}.

\bibitem[BFK{\etalchar{+}}]{ginac}
Christian Bauer, Alexander Frink, Alexander~V. Kisil, Christian Kreckel, Alexei
  Sheplyakov, and Jens Vollinga.
\newblock {GiNaC is Not a CAS}.
\newblock \url{http://www.ginac.de}.

\bibitem[Cla]{clang}
{Clang: A C Language Family Frontend for LLVM}.
\newblock \url{http://clang.llvm.org}.

\bibitem[JU09]{tmlqcd}
Karl Jansen and Carsten Urbach.
\newblock {tmLQCD: A Program Suite to Simulate Wilson Twisted Mass Lattice
  QCD}.
\newblock {\em Comput. Phys. Commun.}, 180:2717--2738, 2009.

\bibitem[KAC{\etalchar{+}}96]{pips}
Ronan Keryell, Corinne Ancourt, Fabien Coelho, B{\'e}atrice Creusillet,
  Francois Irigoin, and Pierre Jouvelot.
\newblock {PIPS: a Workbench for Building Interprocedural Parallelizers,
  Comilers and Optimizers}.
\newblock Technical Paper A/289/CRI, {ENS des Mines Paris}, May 1996.
\newblock \url{http://pips4u.org}.

\bibitem[L{\"u}s05]{ddhmc}
Martin L{\"u}scher.
\newblock {Schwarz-Preconditioned HMC Algorithm for Two-Flavour Lattice QCD}.
\newblock {\em Comput. Phys. Commun.}, 165:199--220, 2005.
\newblock \url{http://luscher.web.cern.ch/luscher/DD-HMC/index.html}.

\bibitem[vG93]{pamela}
Arjan J.~C. van Gemund.
\newblock {Performance Prediction of Parallel Processing Systems: The PAMELA
  Methodology}.
\newblock In {\em Proc. 7th ACM Int. Conf. on Supercomputing}, ICS '93, pages
  318--327, New York, USA, 1993. ACM.

\end{thebibliography}

\end{document}